\documentclass[showpacs,preprint,amsmath,amssymb,nofootinbib,superscriptaddress]{revtex4}

\usepackage{color}
\usepackage{graphicx}
\usepackage{dcolumn}
\usepackage{bm}
\usepackage{lscape}

\begin{document}
\title{Medium effect in high-density region probed by nucleus-nucleus elastic scattering}

\author{T.~Furumoto}
\email{furumoto@ichinoseki.ac.jp}
\affiliation{National Institute of Technology, Ichinoseki College, Ichinoseki, Iwate 021-8511, Japan}

\author{Y.~Sakuragi}%
\affiliation{Department of Physics, Osaka City University, Osaka 558-8585, Japan}

\author{Y.~Yamamoto}
\affiliation{RIKEN Nishina Center, RIKEN, Wako, Saitama 351-0198, Japan}

\date{\today}

\begin{abstract}
We investigate the sensitivity of the medium effect in the high-density region on the nucleus-nucleus elastic scattering in the framework of the double-folding (DF) model with the complex $G$-matrix interaction.
The medium effect including three-body-force (TBF) effect is investigated with two methods.
In the both methods, the medium effect is clearly seen on the potential and the elastic cross section.
Finally, we make clear the crucial role of the TBF effect up to $k_F =$ 1.6 fm$^{-1}$ in the nucleus-nucleus elastic scattering.
\end{abstract}

\pacs{21.30.-x, 24.10.-i, 24.10.Ht, 25.70.Bc} 
\keywords{complex G-matrix, double-folding model, frozen-density approximation}

\maketitle

The optical model potential (OMP) is one of key issues for the nuclear physics not only to analyze the nuclear reaction data but also to understand nuclear reaction mechanism and fundamental interactions between complex nuclear systems.
Historically, the OMP is constructed from the phenomenological way to reproduce the experimental elastic scattering data.
However, the phenomenological OMP, particularly for nucleus-nucleus systems, has a large ambiguity for the inner part of the potential.
Namely, a phenomenological OMP may reproduce the data but it is not always unique.
Then, the more reliable method to construct the OMP is required.
In order to solve the problem, the folding model is proposed.
The model is the simplest but powerful tool to construct the nucleus-nucleus (or nucleon-nucleus) potential by folding the densities of the colliding nuclei with the effective nucleon-nucleon ($NN$) interaction.
For the nucleus-nucleus system, the density-independent $NN$ interaction (M3Y) was first applied with ``success'' to the double-folding (DF) model in the embryonic stage~\cite{M3Y77,SAT79}.
Thereafter, the reliability of the DF potential has been consolidated by the proposed phenomenological density-dependent $NN$ interactions~\cite{KOB82,FARID85,KHO94,BRA97,KHO97}.
Then, the analyses of the nuclear reaction data are advanced with the DF model.
However, those phenomenological density-dependent $NN$ interactions have no imaginary part.
Consequently, the imaginary part of the OMP is still constructed by phenomenological way.

To construct not only the real part but also the imaginary part from the microscopic view point, the complex $G$-matrix interactions have been applied to the DF model~\cite{NAG85,CAR96,TRA00,BLA05,FUR06}.
Those complex $G$-matrix interactions are successful to describe the nucleon-nucleus scattering in the framework of the single-folding model.
The application to nucleus-nucleus system seems like working well and the various experimental data are analyzed.
However, the success of the complex potential is limited, and it is suggested that the existing complex $G$-matrix interactions lack the information in the high-density region (up to twice the normal density) for the nucleus-nucleus system~\cite{FUR06}.
It is pointed out~\cite{FUR06} that the prescription for evaluating the local density as the arithmetic or geometric average of densities of colliding nuclei, such as those adopted in the JLM model, is not suitable for proper estimation of the medium effects in the DF-model calculations.
The OMP is designed to give a potential that reproduces the wave function for the relative motion in the elastic channel where colliding nuclei are kept staying in their ground states~\cite{DNR}.
Hence, it is reasonable to assume that the local density to be used in calculating the DF-model  potential is the sum of the undisturbed densities of the colliding nuclei even at a short relative distances (frozen-density approximation; FDA).
Then, FDA is considered as the most suitable prescription in the DF model and the effect and justification are discussed in Refs.~\cite{FUR06,KHO07,FUR09}.
In the situation, the complex $G$-matrix interaction is needed up to twice the normal density because the local density reaches up to twice the normal density by FDA.

Recently, the present authors propose the complex $G$-matrix interaction, CEG07~\cite{CEG07,FUR09}, whose density dependence is calculated up to twice the normal density ($k_F = 1.8$ fm$^{-1}$).
Then, they have applied the CEG07 interaction to the DF model, and the DF potential with the CEG07 interaction well reproduces the experimental data of the nucleus-nucleus elastic scattering~\cite{FUR09R,FUR09,FUR12}.
The reliable complex optical potential between colliding nuclei is constructed from the microscopic view point.
In Refs.~\cite{FUR09R,FUR09}, it is made clear that the three-body-forces (TBFs) effect, especially the repulsive component of the TBF effect, plays an important role to reproduce the precise angular distribution of the nucleus-nucleus scattering up to backward angles while the TBF effect is introduced phenomenologically.
In Refs.~\cite{CEG07,FUR09,FUR09R}, the three-body repulsive effect was represented conveniently by changing masses of exchanged vector mesons in nuclear medium. 
In Ref.~\cite{YAM13R}, the similar effect was given more clearly as a multi-pomeron exchange potential (MPP).
The strengths of MPP were determined by the analysis of the $^{16}$O~+~$^{16}$O elastic scattering with the DF potential from complex $G$-matrix interactions in the same way as the present work.
The MPP model includes triple and quartic pomeron exchanges, and can lead to the neutron-star EOS stiff enough to reproduce a maximum star mass over $2M_{\odot}$. 
Recently, such a conclusion has been obtained even in hyperon-mixed neutron-star matter \cite{YAM14}.
Thus, our DF model analyses based on FDA are demonstrated to be quite useful to find many-body repulsive effects in high density EOS, for which high-energy central heavy-ion collisions or any other experiments still remain inconclusive.
One of purposes in this work is to make clear the reason why our DF model analyses are so useful to obtain valuable information in the high-density region.

In this Rapid Communication, we apply the complex $G$-matrix interaction based on the ESC08 interaction with the MPP model~\cite{ESC08-1,ESC08-2,YAM13R,YAM14} to the DF model calculation.
In the DF model, it is considered that the local density through the FDA reaches up to about the twice the normal density.
In Refs.~\cite{FUR09R,FUR09}, the important role of the three-body forces effect, especially its repulsive effect, in the nucleus-nucleus scattering was clearly revealed by the DF model with the complex $G$-matrix interaction.
However, the detail of the medium effect in the high-density region beyond the normal density has been left unknown. 
We here clarify the decisive role of the medium effect, including the TBF effect, in such high-density region on the DF potential and scattering observables of the nucleus-nucleus system.
To this end, we demonstrate the importance of properly evaluating the medium effect at high-density region to a sufficient convergence of the calculated DF potential in the spatial region that can be proved by the observed elastic scattering cross sections.

\vspace{2mm}
We construct the nucleus-nucleus potential based on the DF model with the use of the complex $G$-matrix interaction including the TBF effect based on the MPP model. 
The microscopic nucleus-nucleus potential can be written as a Hartree-Fock type potential; 
\begin{eqnarray}
U&=&\sum_{i\in A_{1},j\in A_{2}}{[<ij|v_{D}|ij>+<ij|v_{\rm{EX}}|ji>]}\\
&=&U_{D}+U_{\rm{EX}},
\end{eqnarray}
where $v_{D}$ and $v_{\rm{EX}}$ are the direct and exchange parts of complex $G$-matrix interaction. 
The exchange part is a nonlocal potential in general. 
However, by the plane-wave representation 
for the $NN$ relative motion \cite{SIN75, SIN79}, the exchange part can be localized. 
The direct and exchange parts of the localized potential are then written in the standard form of the DF potential as 
\begin{equation}
U_{D}(R)=\int{\rho_{1}(\bm{r}_1) \rho_{2}(\bm{r}_2) v_{D}(\bm{s}; \rho, E/A)d\bm{r}_1 d\bm{r}_2}, \label{eq:dfdirect}
\end{equation}
where $\bm{s}=\bm{r}_2-\bm{r}_1+\bm{R}$, and
\begin{eqnarray}
U_{\rm{EX}}(R)&=&\int{\rho_{1}(\bm{r}_1, \bm{r}_1+\bm{s}) \rho_{2}(\bm{r}_2, \bm{r}_2-\bm{s}) v_{\rm{EX}}(\bm{s}; \rho, E/A)} \nonumber \\
&&\times \exp{ \left[ \frac{i\bm{k}(R)\cdot \bm{s}}{M} \right] } d\bm{r}_1 d\bm{r}_2. 
\label{eq:dfexchange}
\end{eqnarray}
Here, $\bm{k}(R)$ is the local momentum for nucleus-nucleus relative motion defined by 
\begin{equation}
k^2(R)=\frac{2mM}{\hbar^2} \left\{ E_{\rm{c.m.}}-{\rm{Re}}\ U(R)-V_{\rm{Coul.}}(R) \right\}, \label{eq:kkk}
\end{equation}
where $M=A_1A_2/(A_1+A_2)$, $E_{\rm{c.m.}}$ is the center-of-mass energy, 
$E/A$ is the incident energy per nucleon, $m$ is the nucleon mass and $V_{\rm{Coul.}}$ is the Coulomb potential. 
$A_{1}$ and $A_{2}$ are the mass numbers of the projectile and target, respectively. 
The exchange part is calculated self-consistently on the basis of the local energy approximation through Eq.~(\ref{eq:kkk}). 
Here, the Coulomb potential $V_{\rm{Coul.}}$ is also obtained by folding the $NN$ Coulomb potential with the proton density distributions of the projectile and target nuclei. 
The density matrix $\rho(\bm{r}, \bm{r}')$ is approximated in the same manner as in \cite{NEG72}; 
\begin{equation}
\rho (\bm{r}, \bm{r}')
=\frac{3}{k^{\rm{eff}}_{F}\cdot s}j_{1}(k^{\rm{eff}}_{F}\cdot s)\rho \Big(\frac{\bm{r}+\bm{r}'}{2}\Big), 
\label{eq:exchden}
\end{equation}
where $k^{\rm{eff}}_{F}$ is the effective Fermi momentum \cite{CAM78} defined by
\begin{equation}
k^{\rm{eff}}_{F} 
=\left\{ \left(\frac{3\pi^2}{2} \rho \right)^{2/3}+\frac{5C_{s}(\nabla\rho)^2}{3\rho^2}
+\frac{5\nabla ^2\rho}{36\rho} \right\}^{1/2}, \;\; 
\label{eq:kf}
\end{equation}
where we adopt $C_{s} = 1/4$ following Ref.~\cite{KHO01}. 
The exponential function in Eq.~(\ref{eq:dfexchange}) is approximated by the leading term of the multipole expansion, namely the spherical Bessel function of rank 0, $j_{0} (\frac{k(R)s}{M})$, following the standard prescription~\cite{BRI77,ROO77,BRI78,CEG83,CEG07,MIN10}.

In the present calculations, we employ the FDA for the local density as mentioned in introduction.
In the FDA, the density-dependent $NN$ interaction is assumed to feel the local density defined as the sum of densities of colliding nuclei evaluated;
\begin{equation}
\rho = \rho_{1} (\bm{r}_1) + \rho_{2} (\bm{r}_2). \label{eq:fda}
\end{equation}
The FDA has been widely used also in the standard DF model calculations~\cite{SAT79, KHO94, KHO97, KHO01, KAT02}.
In Ref.~\cite{FUR09}, it is confirmed that FDA is the best prescription in the case with complex $G$-matrix interaction to reproduce the data. 

\vspace{2mm}
We now apply the complex $G$-matrix interactions, which is constructed from the ESC08 interaction with the MPP model~\cite{ESC08-1,ESC08-2,YAM13R}, to nucleus-nucleus systems through the DF model. 
There have been proposed three versions of the MPP model (MPa/b/c)~\cite{YAM14}, which reproduce the $^{16}$O~+~$^{16}$O angular distribution equally well but give rise to different stiffness of EOS.
We use here the MPa version giving the stiffest EOS.
We test the medium effect in high-density region for the elastic scattering of the $^{16}$O~+~$^{16}$O system at $E/A =$ 70 MeV the same as in Refs.~\cite{FUR09R,FUR09}.
Because the $^{16}$O nucleus is one of the most stable double magic nuclei and has no collective excited states strongly coupled to the ground state, the $^{16}$O~+~$^{16}$O system is considered as an ideal system to probe the potential at short distances that directly reflects the medium effect in the high-density region.
We adopt the nucleon density of the $^{16}$O nucleus calculated from the internal wave functions generated by the orthogonal condition model (OCM) by Okabe~\cite{OKABE} based on the microscopic $\alpha$~+~$^{12}$C cluster picture.

First, we investigate the medium effect for the high-density region in the framework of the DF model with complex $G$-matrix interaction.
We already mentioned the importance of the medium effect, especially the TBF effect, for the nucleus-nucleus elastic scattering in Refs.~\cite{FUR09R,FUR09}.
However, the detail of the medium effect in the high-density region is not investigated.
Then, we test the sensitivity of the medium effect in the high-density region by the following artificial cut of the evaluated local density; 
\begin{eqnarray}
\rho&=&
\left\{
  \begin{array}{cccc}
   \rho_{1}+\rho_{2} & \ldots & (\text{if \ \ }\rho_{1}+\rho_{2} < \rho_{\rm cut} ) & \\
   \rho_{\rm cut} & \ldots & (\text{if \ \ }\rho_{1}+\rho_{2} > \rho_{\rm cut} ) & ,
  \end{array}
\right. \label{eq:cut}
\end{eqnarray}
where the $\rho_{\rm cut}$ value is varied as a parameter.
We calculate the DF potentials with several $k_{\rm cut}$ values where $k_{\rm cut}$ is defined by
\begin{equation}
\rho_{\rm cut} = \frac{2}{3\pi^2}k_{\rm cut}^3. \label{eq:krho}
\end{equation}
By changing the $k_{\rm cut}$ value, the medium effect in the high-density region is controlled and investigated in the potential and observable cross section.

\begin{figure}[h]
\begin{center}
\includegraphics[width=6.5cm]{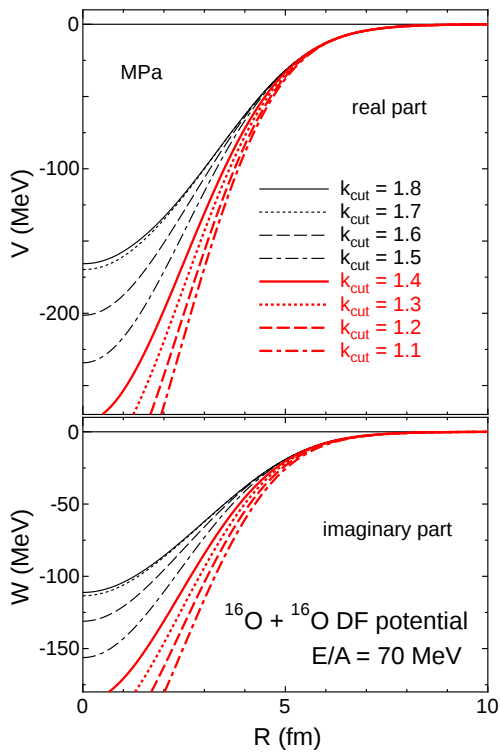}
\caption{\label{fig:01} (Color online) The real and imaginary parts of the DF potential with the $k_{\rm cut}$ value. 
The solid, dotted, dashed, dot-dashed, bold-solid, bold-dotted, bold-dashed, and bold-dot-dashed curves are the results with $k_{\rm cut} = 1.8, 1.7, 1.6, 1.5, 1.4, 1.3, 1.2,$ and $1.1$, respectively.}
\end{center}
\end{figure}
Figure~\ref{fig:01} shows the real and imaginary parts of the calculated DF potential for the $^{16}$O~+~$^{16}$O elastic scattering at $E/A$ = 70 MeV.
The medium effect is clearly seen for the complex potential, especially for the inner part of the potential.
For the inner part, the local density based on the FDA reaches up to twice the normal density by approaching each other nucleus.
On the other hand, the complex $G$-matrix interaction almost feels small density in the tail region, and then, the change by the $k_{\rm cut}$ value around the tail part of the potential is small.
When the local density is restricted by Eq.~(\ref{eq:cut}), the complex $G$-matrix interaction feels the restricted medium effect and gives the strong potential.
Therefore, the DF model restricted by several $k_{\rm cut}$ values gives the deep potential.
In the potential, the medium effect is clearly seen up to $k_F =$ 1.8 fm$^{-1}$ (solid curve).
Namely, it is indicated that the medium effect in the high-density region up to $k_F =$ 1.8 fm$^{-1}$ has an important role to construct the DF potential.
By way of caution, we here mention that the local density can not reach over the $k_F =$ 1.8 fm$^{-1}$ (twice the normal density) in principle because the local density is composed of the sum of the density of the colliding two nuclei as Eq.~(\ref{eq:fda}).
In addition, we disclose the imperfection of the DF potential with the JLM~\cite{JLM77}, original CEG~\cite{CEG83,CEG86}, and Melbourne-$G$~\cite{MEL00} interactions, whose medium effect are not calculated up to twice the normal density.

\begin{figure}[h]
\begin{center}
\includegraphics[width=7cm]{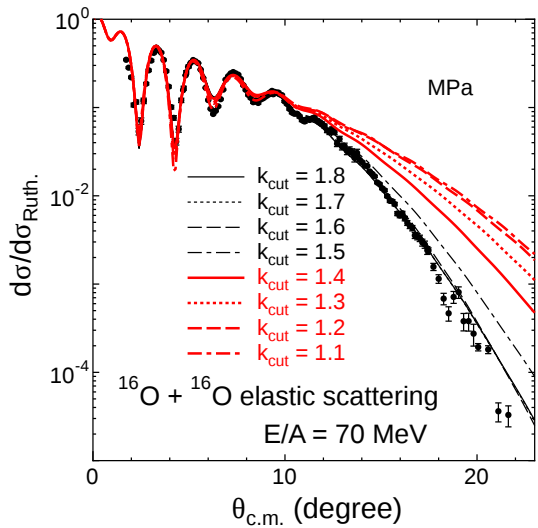}
\caption{\label{fig:02} (Color online) The elastic cross section with the DF potentials shown in Fig.~\ref{fig:01}. 
The meaning of the curves is the same as in Fig.~\ref{fig:01}.
The experimental data is taken from Ref.~\cite{NUO98}}
\end{center} 
\end{figure} 
Figure~\ref{fig:02} shows the results calculated with the DF potentials shown in Fig.~\ref{fig:01} for the $^{16}$O~+~$^{16}$O elastic scattering at $E/A =$ 70 MeV.
The medium effect is clearly seen up to $k_F =$ 1.6 fm$^{-1}$ in the elastic cross section while the effect is seen up to $k_F =$ 1.8 fm$^{-1}$ in the potential.
The difference of the $k_F$ values comes from the insensitivity of the elastic scattering to the most inner part of the potential.
These results clearly show the importance of the proper evaluation of the medium effect in the high-density region ($k_F >1.4$--$1.8$ fm$^{-1}$ ) and raise a strong caution to apply the existing complex $G$-matrix interaction to the analyses of heavy-ion scattering through the DF model calculations.
In other words, the present result implies that the medium effect in the high-density region can be probed rather sensitively through the nucleus-nucleus elastic scattering experiments.
We also investigate the medium effect in the high-density region on the total reaction cross section but the effect is found to be negligible.

\begin{figure}[b]
\begin{center}
\includegraphics[width=6.5cm]{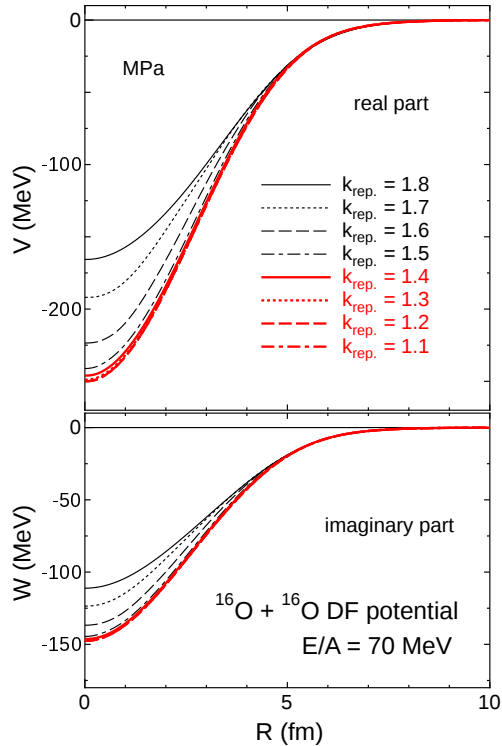}
\caption{\label{fig:03} (Color online) The real and imaginary parts of the DF potential with the $k_{\rm rep.}$ value. 
The solid, dotted, dashed, dot-dashed, bold-solid, bold-dotted, bold-dashed, and bold-dot-dashed curves are the results with $k_{\rm rep.} = 1.8, 1.7, 1.6, 1.5, 1.4, 1.3, 1.2,$ and $1.1$, respectively.}
\end{center} 
\end{figure}
\begin{figure}[t]
\begin{center}
\includegraphics[width=7cm]{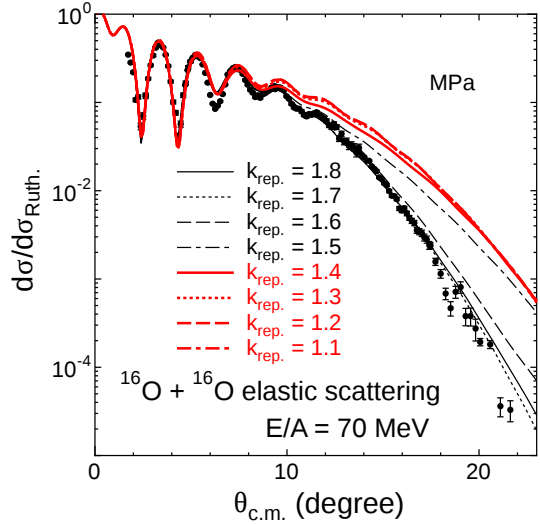}
\caption{\label{fig:04} (Color online) The elastic cross section with the DF potentials shown in Fig.~\ref{fig:03}. 
The meaning of the curves is the same as in Fig.~\ref{fig:03}.}
\end{center} 
\end{figure} 

Next, we focus the role of the TBF effect in the high-density region.
In Refs.~\cite{FUR09R,FUR09}, we make clear the important role of the TBF effect for the nucleus-nucleus elastic scattering.
We here test the sensitivity of the TBF effect in the high-density region by the following prescription for the complex $G$-matrix interaction; 
\begin{eqnarray}
v(\bm{s}; \rho, E/A)&=&
\left\{
  \begin{array}{rl}
{\rm MPa} & {\rm (with\ TBF)} \\
\ldots & (\text{if \ \ } \rho = \rho_{1}+\rho_{2} < \rho_{\rm rep.} ) \\
{\rm ESC} & {\rm (w/o\ TBF)} \\
\ldots & (\text{if \ \ } \rho = \rho_{1}+\rho_{2} > \rho_{\rm rep.} ) \ ,
  \end{array}
\right. \label{eq:rep}
\end{eqnarray}
where the ESC interaction is the complex $G$-matrix interaction constructed only from the ESC08 interaction.
Namely, the ESC interaction does not include the TBF effect.
The $\rho_{\rm rep.}$ is the parameter.
We calculate the DF potentials with several $k_{\rm rep.}$ values where $k_{\rm rep.}$ is defined by
\begin{equation}
\rho_{\rm rep.} = \frac{2}{3\pi^2}k_{\rm rep.}^3. \label{eq:krho2}
\end{equation}
Namely, we replace MPa (with TBF) by ESC (w/o TBF) when the local density $\rho$ exceeds the $\rho_{\rm rep.}$ value.
By changing the $k_{\rm rep.}$ value, the TBF effect in the high-density region is investigated in the potential and observable cross section.

Figure~\ref{fig:03} shows the real and imaginary parts of the calculated DF potential for the $^{16}$O~+~$^{16}$O elastic scattering at $E/A$ = 70 MeV.
The TBF effect is clearly seen for the complex potential, especially for the inner part of the potential.
In the potential, the TBF effect is clearly seen up to $k_F =$ 1.8 fm$^{-1}$.
This result indicates the importance of the TBF effect in the high-density region.
Here, we notice that the difference between the solid curve and $k_{\rm rep.} =$ 1.7 fm$^{-1}$ in Fig.~\ref{fig:03} is larger than that between the solid curve and $k_{\rm cut} =$ 1.7 fm$^{-1}$ in Fig.~\ref{fig:01}.
Namely, the potential with $k_{\rm rep.} =$ 1.7 fm$^{-1}$ in Fig.~\ref{fig:03} gives deeper potential than that with $k_{\rm cut} =$ 1.7 fm$^{-1}$ in Fig~\ref{fig:01}.
This cause comes from the strength of the MPa interaction and that of the ESC interaction at $k_{\rm rep.} =$ 1.7--1.8 fm$^{-1}$.
In fact, the single-particle potential ($U$) in the nuclear matter for the several cases is obtained as Re $U_{\rm MPa} (k_F = $1.7 fm$^{-1}) = -30.60$ MeV, Re $U_{\rm MPa} (k_F = 1.8$ fm$^{-1}) = -16.37$ MeV, Re $U_{\rm ESC} (k_F = 1.7$ fm$^{-1}) = -64.58$ MeV, and Re $U_{\rm ESC} (k_F = 1.8$ fm$^{-1}) = -67.76$ MeV.
The potential of $k_{\rm rep.} =$ 1.7 fm$^{-1}$ in Fig.~\ref{fig:03} gives deeper potential than that of $k_{\rm cut} =$ 1.7 fm$^{-1}$ in Fig.~\ref{fig:01} because Re $U_{\rm ESC} (k_F = $1.8 fm$^{-1})$ gives more attractive potential than Re $U_{\rm MPa} (k_F = $1.7 fm$^{-1})$.

Figure~\ref{fig:04} shows the results calculated with the DF potentials shown in Fig.~\ref{fig:03} for the $^{16}$O~+~$^{16}$O elastic scattering at $E/A =$ 70 MeV, respectively.
The TBF effect is clearly seen up to $k_F =$ 1.6 fm$^{-1}$ in the elastic cross section while the TBF effect up to $k_F = 1.8$ fm$^{-1}$ is seen in the DF potential.
Here, it is first survey that the information of the medium effect in such high-density region (up to $k_F =$ 1.6 fm$^{-1}$) is extracted from the observation.
The important role of the TBF effect in the high-density region is confirmed in the nucleus-nucleus elastic cross section.
This result again implies that the nucleus-nucleus elastic scattering can sensitively probe the important role of the TBF effect in the high-density region up to $k_F = 1.6$ fm$^{-1}$.

\vspace{2mm}
In summary, we have constructed the DF potential with the complex $G$-matrix interaction including the TBF effect based on the MPP model.
With the DF potential, the medium effect in the high-density region have been investigated by cutting the local density in the high-density region.
The medium effect is clearly seen in the potential and the elastic scattering cross section.
In addition, the TBF effect in the high-density region have also been investigated.
The TBF effect up to $k_F =$ 1.6 fm$^{-1}$ has a critical role to determine the elastic angular distribution.
Here, we should mention that similar results are obtained with the use of the CEG07 or MPb/c interactions~\cite{CEG07,YAM14} instead of the MPa one adopted here and for the other systems.
The result will be presented in the forthcoming paper.
These results imply that the medium effect, especially the TBF effect, in the high-density region can be probed by the observed nucleus-nucleus elastic scattering.
Again, we have raised a caution in the use of the $G$-matrix interactions such as the JLM, the original CEG and Melbourne interactions, which are not arranged the medium effect including TBF effect in the high-density region, to the analyses of nucleus-nucleus scattering/reactions with the DF model potential based on the FDA.
Finally, we made clear the crucial role of the TBF effect in the high-density region on the nucleus-nucleus elastic scattering.

\vspace{2mm}
{\it Acknowledgment. The authors acknowledge Professor Uesaka for encouraging comments.}


\end{document}